\documentclass[12pt]{iopart}
\usepackage{graphicx}
\usepackage{epstopdf,cite}

\begin{document}

\title[Search for single sources of ultra high energy cosmic rays on the sky]{Search for single sources of ultra high energy cosmic rays on the sky}

\author{G. Giacinti$^1$, X. Derkx$^1$, D. V. Semikoz$^{1,2}$}

\address{$^1$ {\it AstroParticle and Cosmology (APC), 10, rue Alice Domon et L\'eonie Duquet, 75205 Paris Cedex 13, France}}
\address{$^2$ {\it Institute for Nuclear Research of the Russian Academy of Sciences, 60th October Anniversary prospect 7a, Moscow 117312, Russia}}

\ead{\mailto{giacinti@apc.univ-paris7.fr}, \mailto{dmitri.semikoz@apc.univ-paris7.fr}}

\begin{abstract}
In this paper, we suggest a new way to identify single bright sources of Ultra High Energy Cosmic Rays (UHECR) on the sky, on top of background. We look for doublets of events at the highest energies, E~$>$~6~$\cdot$~10$^{19}$eV, and identify low energy tails, which are deflected by the Galactic Magnetic Field (GMF). For the sources which are detected, we can recover their angular positions on the sky within one degree from the real ones in 68\% of cases. The reconstruction of the deflection power of the regular GMF is strongly affected by the value of the turbulent GMF. For typical values of 4$\mu$G near the Earth, one can reconstruct the deflection power with 25\% precision in 68\% of cases.
\end{abstract}





\maketitle

\section{Introduction}

The search for the Ultra-High Energy Cosmic Ray
(UHECR) sources is one of the most interesting problems of modern
astroparticle physics. The observation of the Greisen-Zatsepin-Kuzmin (GZK) cutoff~\cite{GZK} in the energy
spectrum by HiRes experiment~\cite{spectrum_HiRes}
proves the astrophysical origin of UHECR.

Several classes of astrophysical sources can potentially accelerate cosmic rays
to the highest energies~\cite{Hillas:1985is}, but some of them are restricted
by different energy losses in the sources (for a recent review see, for example,
Ref.~\cite{Ptitsyna:2008zs}). Also, for each class of sources,
several alternative acceleration mechanisms were suggested, including standard first
order Fermi acceleration in shocks or acceleration in induced electric fields in the
polar regions of supermassive black holes~\cite{Neronov:2007mh}.

Accelerated cosmic rays lose their energy during the propagation from the sources
to the Earth due to pion production, e$^{\pm}$ pair production and redshifting.
These energy losses define the horizon - the typical distance from which cosmic rays
can reach us with a final energy bigger than some value. The horizon value was discussed in Ref.~\cite{Harari:2006uy}, in the constant energy loss approximation. It was updated to realistic stochastic energy losses in Ref.~\cite{Kalashev:2007ph}. Finally, the effects of the experimental energy resolution were taken into account in~\cite{Kachelriess:2007bp}. Both for protons and iron nuclei, 70\% of sources would be located within 70~Mpc of the Earth for energies E~$>$~10$^{20}$~eV, and within 250~Mpc of the Earth for E~$>$~6~$\cdot$~10$^{19}$~eV.

Since the Large Scale Structure (LSS) is still anisotropic on 100-200~Mpc scales, any class of astrophysical sources of UHECR should give an anisotropic signal in detectors
with typical angular scales of tens of degrees. Anisotropies on medium scales, 20-30
degrees, have been found previously by combining all the available data
of ``old'' cosmic ray experiments~\cite{msc}. Such anisotropies were
predicted as a consequence of the observed LSS of
matter~\cite{old} and favor therefore, with the presence
of the Greisen-Zatsepin-Kuzmin cutoff, the extragalactic origin
of UHECRs.

These results still did not allow to find the sources of UHECR, both due to limited statistics
and to the fact that any different class of astrophysical sources is located
in the same LSS, which makes their identification more difficult.

Another important factor which complicates the detection of cosmic ray sources is the deflection of UHECR in the galactic and extragalactic magnetic fields.
Extragalactic fields are unknown outside of galaxy clusters. The numerical simulations done
by several groups predict a wide range of deflections for 10$^{20}$~eV protons. According to Ref.~\cite{Dolag:2004kp}, deflections are negligible ($\delta<$~1$^{\circ}$
in 98\% of the sky within 100~Mpc of the Earth). On the contrary, according to Ref.~\cite{Sigl:2004yk}, deflections are more than 10~degrees in 70\% of the sky. In this case, the search
for the UHECR sources is extremely difficult and close to impossible, even for proton primaries.
In the following, we will assume that the deflections in the extragalactic
magnetic fields are small, following the results of Ref.~\cite{Dolag:2004kp}.
Let us note that if the sources are located in regions with high values of magnetic fields,
like galaxy clusters, both their spectrum and their detected flux can be affected by
magnetic lensing effects in the cluster~\cite{Dolag:2008py}. Below, we assume that one can neglect such effects.

Even in case the extragalactic magnetic fields are negligible, UHECR would still be deflected 
in the Galactic Magnetic Field (GMF). For the GMF, the situation is much better due to existing measurements of
both the regular and the turbulent components of this field. However, there is no theoretical model of the GMF which can fit all existing data. The measured values of the GMF still allow to estimate the typical deflections of UHECR in it, which are $\simeq$~0.5-2~degrees for protons of energies equal to 10$^{20}$~eV, depending on the position on the sky and on the GMF model.

There are two possible ways to continue the search for the UHECR sources in the near future.
One is to look at the highest energies only, E~$>$~6~$\cdot$~10$^{19}$~eV, and collect
enough statistics to find point sources. This method can be strongly affected
by the presence of a large fraction of nuclei in the UHECR flux.

In this paper, we present another way to search for individual bright sources.
One can hope that the regular GMF gives the major contribution to UHECR
deflections. Then, one can try to look near high energy events for low energy tails on top of a
high background of UHECR, which comes from all unresolved proton sources and from heavy nuclei.
In the present paper, we suggest a new method which is based
on searching for low energy tails near doublets or clusters of events with high energies,
E~$>$~10$^{19.8}$~eV $\simeq$~6.3~$\cdot$~10$^{19}$~eV. This allows us to reduce the 
background significantly, compared to usual searches for tails near every single high energy event.

We simulated the signals from a single source and from a background of UHECR. Following the results of HiRes~\cite{BelzICRC}, we take a proton source, except in a paragraph of Section~\ref{results}, where we investigate some possible extensions of this work to sources of heavier nuclei.

Then, we studied how the probability to detect this source depends on unknown parameters of the source and of the magnetic field. In case the source is detected, we finally reconstruct its position on the sky and the local deflection power of the regular GMF.

In Section~\ref{simulation}, we simulate the signal from a single source on top of background. Section~\ref{sectionmethod} depicts the method used to detect the source and reconstruct its position. In Section~\ref{results}, we test how its efficiency varies with experimental or unknown physical parameters. In Section~\ref{summary}, we summarize our results.

\section{Simulation of the signal from a single source on top of background}
\label{simulation}

For this study, we only take into account the cosmic rays which energies are bigger than a given energy threshold, E$_{th}$. Typically, E$_{th}$~$\geq$~10$^{19}$~eV. 
Detecting a signal from a source on top of background at lower energies is hopeless for realistic magnetic field parameters.

We start below with the simulation of the source signal.

\subsection{Events from the source}
\label{source}

\begin{figure}
\begin{center}
\includegraphics[width=0.48\textwidth]{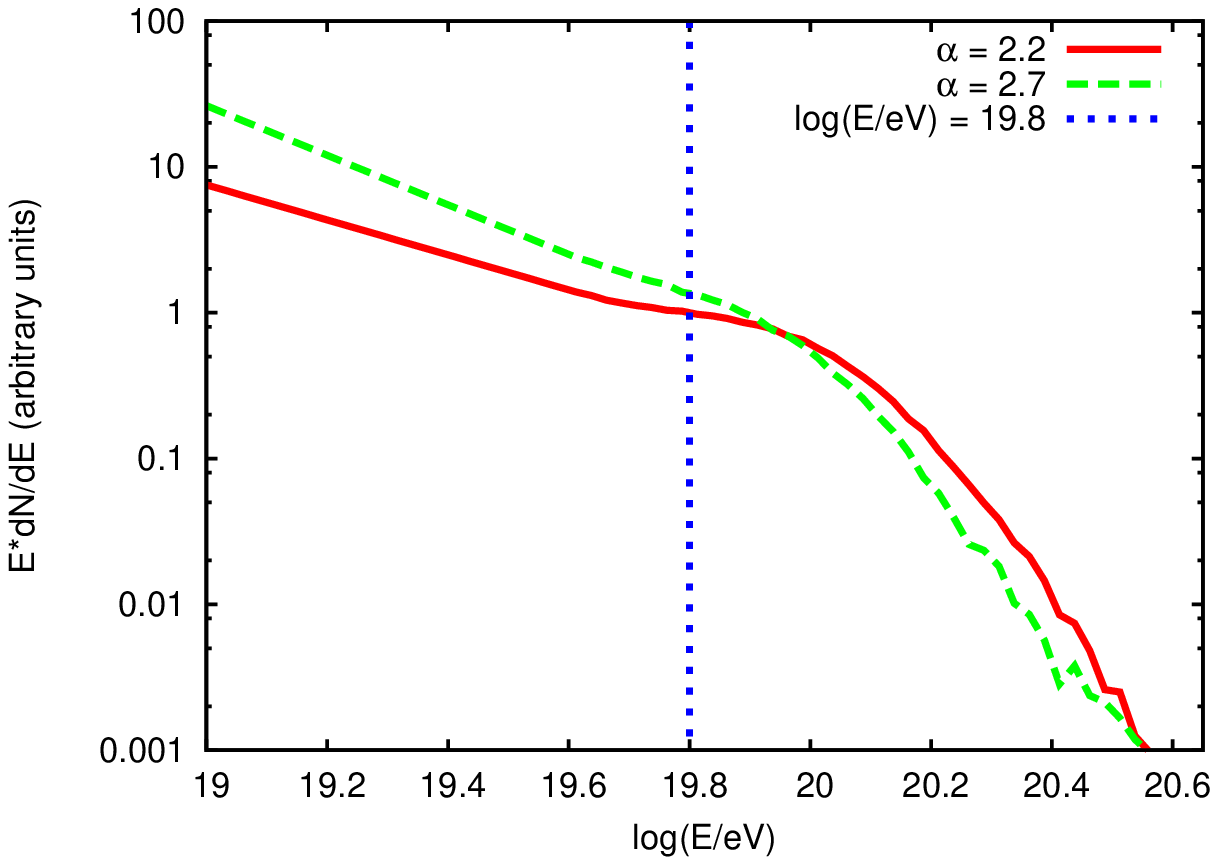}
\includegraphics[width=0.48\textwidth]{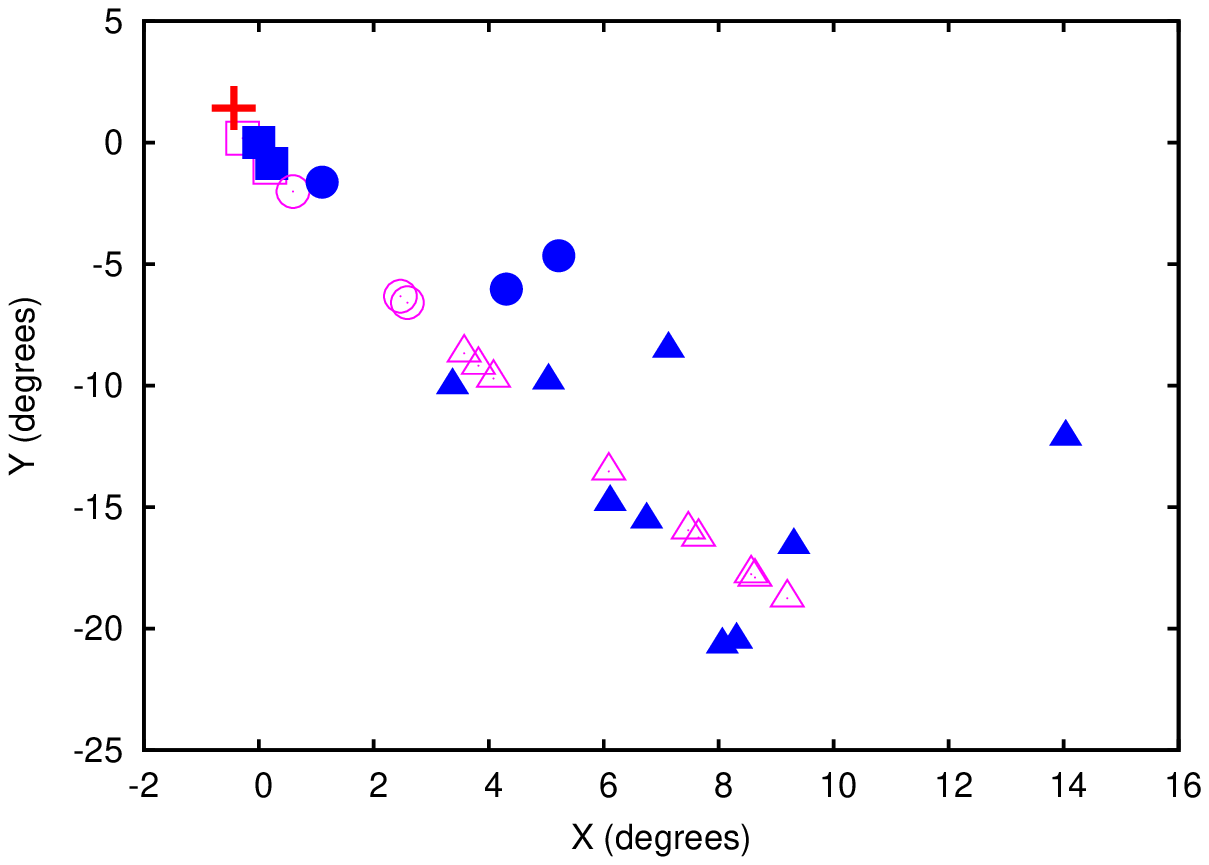}
\end{center}
\caption{Flux (left) and arrival directions (right) of UHECR protons after propagation on 50~Mpc and deflection in the Galactic Magnetic Field. {\bf Left:} solid red line for the power law index $\alpha$~=~2.2 and green dashed line for $\alpha$~=~2.7. Both spectra normalized to the same flux above 10$^{19.8}$~eV. {\bf Right:} projections of the arrival directions of cosmic rays emitted by a source (red cross) in the plane tangent to the celestial sphere and centered on the highest energy event (blue filled square in (0,0)). The deflections X and Y are measured on two orthogonal axes, in degrees. Open symbols for the CR from the source, deflected by the regular GMF only. Filled symbols for the same CR when the turbulent component is also taken into account. The shapes correspond to the CR energies (triangles: below 10$^{19.3}$~eV, circles: between 10$^{19.3}$ and 10$^{19.8}$~eV, squares: above 10$^{19.8}$~eV).}
\label{spectraS}
\end{figure}

We assume that the source has a power law acceleration spectrum in the form:
\begin{equation}
\mbox{F(E)} = f \frac{1}{\mbox{E}^{\alpha}} \Theta \left(\mbox{E}_{max}-\mbox{E}\right), 
\end{equation}
where $f$ is the flux normalization factor, $\alpha$ the power law index, $\Theta$ the Heaviside step function and E$_{max}$~=~10$^{21}$~eV the maximum energy to which the source can accelerate cosmic rays. Below, we consider two examples of power law spectra:  $\alpha$~=~2.2  for Fermi acceleration and $\alpha$~=~2.7 for V.~Berezinsky  et al. model~\cite{Berezinsky:2002nc}.

Since we would like to study a bright nearby proton source, we take typical distances of 50-100~Mpc and
propagate the protons to the Earth. UHECR protons lose energy on their way to the Earth, due to pion production (GZK effect), e$^{\pm}$ pair production and the expansion of the Universe. We use the results of Ref.~\cite{Kachelriess:2007bp} for the energy losses.
Typical spectra are presented in Fig.~\ref{spectraS}a for a source located at 50~Mpc from the Earth.
The two curves are normalized to the same luminosity above E~=~10$^{19.8}$~eV~$\simeq$~60~EeV.
For such small distances, the propagation affects the injection spectrum only at high energies, E~$>$~10$^{19.6}$~eV~$\simeq$~40~EeV, creating the ``bump''~\cite{bump} and the cutoff in the spectrum.

The number of events from the source is computed according to its luminosity. It is defined as the fraction of source events with energies above 10$^{19.8}$~eV in all the sky. In practice, the number of source events which are generated above E$_{th}$ in every Monte Carlo simulation fluctuates around the mean value inferred from the source luminosity.

On the way from the source to the Earth, cosmic rays are deflected in the Galactic Magnetic Field.
This field consists of a regular and a turbulent component which are different in the disk and in the halo.
At high energies, E~$>$~40~EeV, and outside of the galactic plane, the deflections of cosmic rays in the regular field $\delta_{reg}$, can be estimated as (see e.g. Ref.~\cite{Harari2}):

\begin{equation}
\delta_{reg} \simeq 8.1^{\circ} \frac{40 \mbox{EeV}}{\mbox{E/Z}} \left| \int_{0}^{L} \frac{d{\bf s}}{3 \mbox{kpc}} \times \frac{{\bf B}}{2 \mu \mbox{G}} \right|.
\end{equation}
In other words,
\begin{equation}
\delta_{reg} \simeq \frac{\mathcal{D}}{\mbox{E}}~,
\label{1/E}
\end{equation}
  where $\mathcal{D}$ is the deflection power of the regular field for a given region of the sky.

The regular component shifts the cosmic rays as 1/E along a curve. This curve can be approximately straight outside of the galactic plane for high enough energies. Since the magnetic field in the disk is poorly known, one has to exclude the galactic plane for single source searches in any case.

As an example of regular field, we take a model based on the model of Prouza and Smida~(PS)~\cite{GMF,PS}. The PS model consists of a bispiral field in the disk and a halo field containing a poloidal and a torroidal component. We updated this model, according to the latest results of Refs.~\cite{Waelkens:2008gp,Sun:2007mx,Jansson:2009ip,Han:2009jg,Han:2009ts}, and a discussion during the Ringberg workshop~\cite{Ringberg}. For instance, we choose an exponentially decaying profile along the galactocentric radius r, instead of a 1/r profile and we set the value of the regular field close to the Sun, B$_{reg}$, to 2 $\mu$G.

For the turbulent component, we take the model detailed in Refs.~\cite{TurbulentComponent1,Tinyakov:2004pw}. For particles travelling in a turbulent field, the distribution of angular deflections is a two-dimensional Gaussian centered on zero (no deflection). 




 The root mean square deflection due to the turbulent field, $\delta_{rms}$, can be written as:

\begin{equation}
\delta_{rms} = \frac{1}{\sqrt{2}} \frac{\mbox{ZeB}_{rms}}{\mbox{E}} \sqrt{\mbox{LL}_{c}} \simeq 5.8^{\circ} \cdot \left( \frac{10^{19} \mbox{eV}}{\mbox{E/Z}} \right) \left( \frac{\mbox{B}_{rms}}{4 \mu \mbox{G}} \right) \sqrt{\frac{\mbox{L}}{3 \mbox{kpc}}} \sqrt{\frac{\mbox{L}_{c}}{50 \mbox{pc}}}
\label{eqTC}
\end{equation}
where B$_{rms}$ is the RMS value of the turbulent magnetic field, L$_{c}$ its correlation length, and L~$\gg$~L$_{c}$ the length travelled by the cosmic rays in it. We take L$_{c}$~=~50~pc, which is a commonly admitted value~\cite{Ringberg}.

In this work, we take the following profile for B$_{rms}$:

\begin{equation}
\mbox{B}_{rms} = B(r) \exp\left(-\frac{|z|}{z_{0}}\right)~,
\end{equation}
where r is the galactocentric radius, and z the distance to the galactic plane.
We take here $z_{0}=1.5$~kpc.

The radial profile of the field can be defined as:
\begin{equation}
B(r) = \left\{ \begin{array}{ll}
                \mbox{B}_{rms0} \cdot  \exp{\left( \frac{5.5}{8.5} \right) }  & \mbox{, if } \mbox{r} \leq 3 \mbox{kpc (bulge)}\\
		\mbox{B}_{rms0} \cdot \exp{\left( \frac{ - \left( \mbox{r} - 8.5 \mbox{kpc}\right)}{8.5 \mbox{kpc}} \right) } & \mbox{, if }  \mbox{r} > 3 \mbox{kpc}
               \end{array} \right.
\end{equation}
where B$_{rms0}$ denotes the value of B$_{rms}$ close to the Sun. The currently admitted value is B$_{rms0}$~$\simeq$~4~$\mu$G~\cite{Ringberg}. We take it as a reference value, but we will also discuss about the impact of weaker and stronger fields.

One can note that in a given region of the sky, $\delta_{reg}$ and $\delta_{rms}$ are approximately proportional to 1/E. Therefore, for a given energy one has approximately $\delta_{rms} \propto \delta_{reg}$:
\begin{equation}
\delta_{rms} \simeq \mathcal{R}_{turb/reg} \cdot \delta_{reg}
\label{prop}
\end{equation}

$\mathcal{R}_{turb/reg}$ estimates the ratio of deflections respectively due to the turbulent and regular fields.

In terms of deflections, the impact of the two GMF components can be seen in Fig.~\ref{spectraS}b. The regular component shifts the cosmic rays as 1/E along a curve. For each energy, the turbulent component spreads the cosmic rays around the position where they would have been deflected in the regular field only. Due to Eqn.~(\ref{prop}), the cosmic rays are roughly spread in a sector. When there are many events with the same energies, they are not spread uniformly, as discussed in Refs.~\cite{Harari2,TurbulentComponent1,Roulet:2003rr,Golup:2009zg}.
Note that our study does not depend on the details of the Galactic Magnetic Field model. 
The results mostly depend on two parameters that summarize the impact of the regular and turbulent fields. The first of them is the deflection power of the regular field in the region surrounding the source, $\mathcal{D}$ -see Eqn.~(\ref{1/E}). The second one is the proportionality factor $\mathcal{R}_{turb/reg}$ in Eqn.~(\ref{prop}).

One can think that the source detection becomes easier when lowering the energy threshold E$_{th}$, thanks to the increasing number of events. However, it is not so, because the background grows much faster at low energies. We will discuss this in the next section.

\subsection{Background}

\begin{figure}
\includegraphics[width=0.5\textwidth]{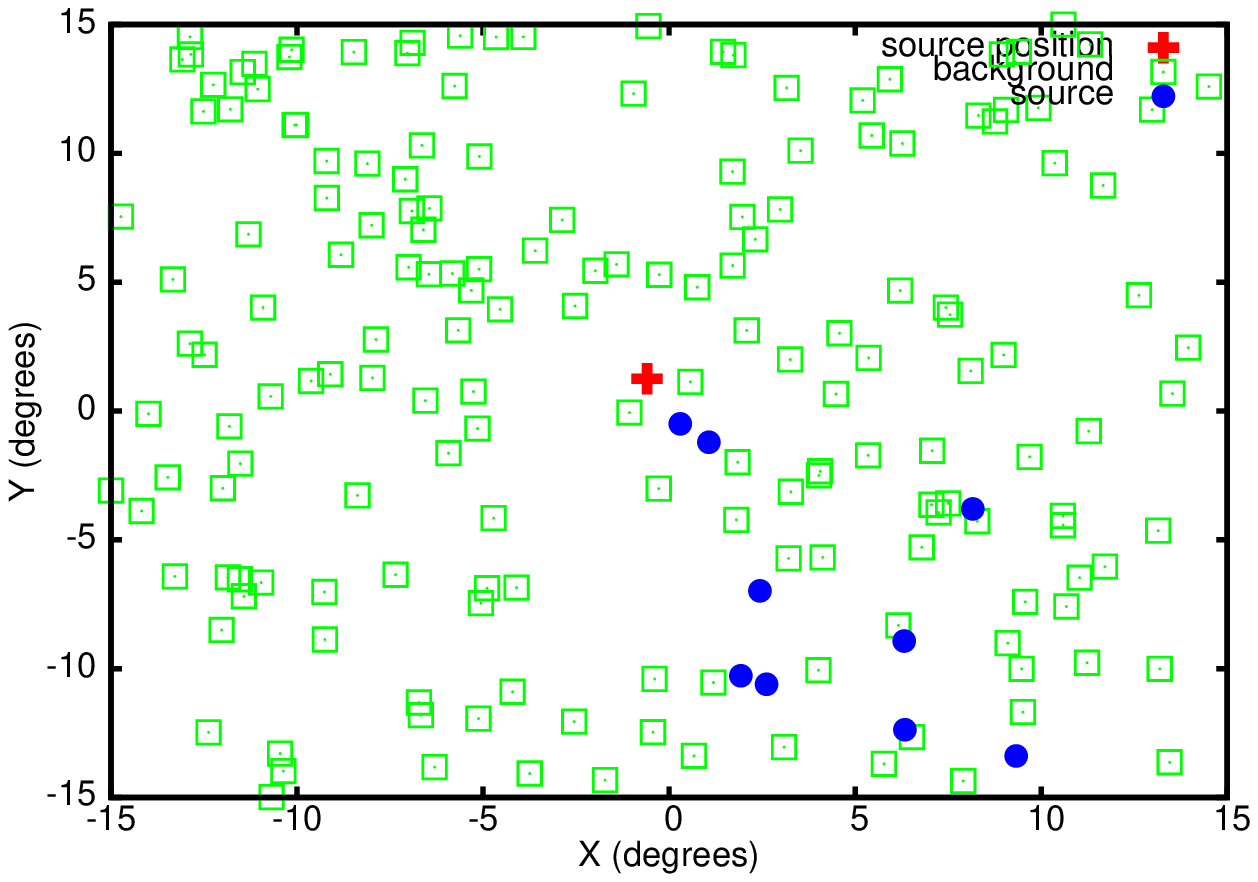}
\includegraphics[width=0.5\textwidth]{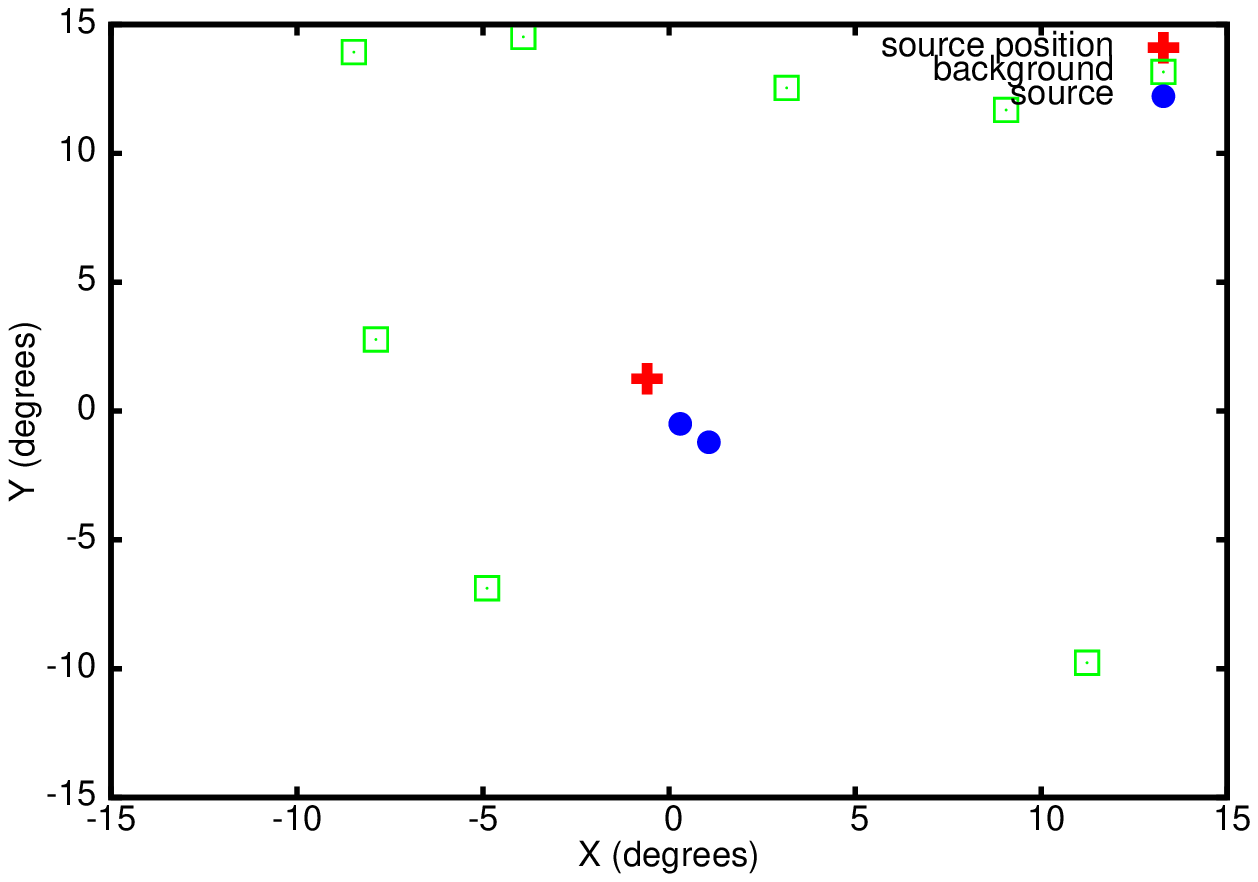}
\caption{Projections of the arrival directions of CR in the tangent plane as in Fig.~\ref{spectraS}b. Red cross for the source position, blue filled circles for source events and green open squares for background events. {\bf Left panel (a):} energy threshold equal to E$_{th}$~=~10$^{19.0}$~eV; {\bf Right panel (b):} same picture, but with a higher energy threshold, equal to E$_{th}$~=~10$^{19.6}$~eV.}
\label{BandS}
\end{figure}

We generate the background according to the exposure of Telescope Array, following Ref.~\cite{exposure}. We set the declination of Telescope Array to $\delta=40.1^{\circ}$ and its maximum zenith angle to $\theta=45^{\circ}$. The energies of background events are generated according to the spectrum measured by HiRes~\cite{spectrum_HiRes}.

Fig.~\ref{BandS} shows source and background events in the region around the source for different energy thresholds. In Fig.~\ref{BandS}a we plot the arrival directions of cosmic rays with energies bigger than E$_{th}$~=~10$^{19.0}$~eV, and in Fig.~\ref{BandS}b, with energies bigger than E$_{th}$~=~10$^{19.6}$~eV. One can clearly notice two important facts for source detection. At high energies, on the one hand, we lose in terms of statistics because we have less cosmic rays from the source, but on the other hand, the background is more strongly reduced than the source signal. This is due to the difference of slopes between the background and the source spectra. That is why there should be an optimum energy threshold E$_{th}$ for source detection between 10$^{19}$~eV and the very highest energies.

For deflections lower than 10 degrees, when one can neglect the dependence on exposure on the considered portion of the sky, our results do not depend on the position of the source because both the signals from the source and the background follow the same exposure. The regions of very small exposure are particular cases. We need at least 3 events from the source for our analysis and in these regions this condition can be difficult to hold for not very bright sources. For the results we discuss in this paper, we put the source in a region where the exposure of Telescope Array is not maximal but still significant.

In Section~\ref{results}, we also investigate separately how our results are affected by the anisotropy at energies above 10$^{19.8}$~eV.

%
%
%
%
%
%

\section{Method to find the source on top of background}
\label{sectionmethod}

In the previous Section, we described how we generate the background events and those coming from the source.
Now, we mix all of them in the same "data" file and try to find the source on top of the background. 
For this purpose, we develop a new source detection method in the next section. When the source is detected, we reconstruct its position on the sky and the deflection power of the regular magnetic field near the source, $\mathcal{D}$.

\subsection{Source detection and background rejection}

\begin{figure}
\centering
\includegraphics[width=0.7\textwidth]{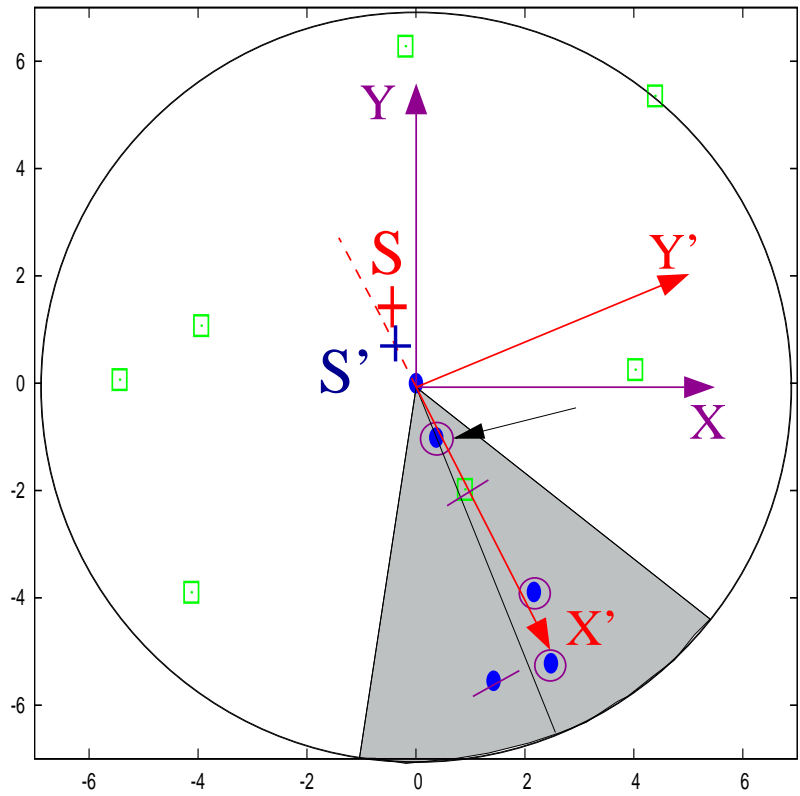}
\caption{Example of source detection on top of background. The big black circle with a radius according to the Eqn.~(\ref{radius}) is centered on the highest energy event, in (0,0). The black arrow points at the second highest energy event. The sector in the direction of this event is highlighted in grey. Blue circles indicate source events. Green squares show positions of background events.  Events that pass the final selection are surrounded with magenta circles. (X,Y) and (X',Y') respectively denote old and new systems of coordinates. S and S' for real and reconstructed source positions, respectively.}
\label{method}
\end{figure}

Usually, one starts with a single high energy event with an energy E$_{1}$ satisfying, for example: E$_{1}$~$\geq$~10$^{19.8}$~eV $\simeq$~6.3~$\cdot$~10$^{19}$~eV. This guarantees two things. First, the source is located nearby in the local Universe, at distances smaller than 250~Mpc in 70\% of cases -see Ref.~\cite{Kachelriess:2007bp}. The source density within such distances should not be too large to prevent the identification of the brightest sources. Second, for standard power law injection spectra with indexes $\alpha$~$>$~2, one should see some low energy events from the source near this high energy event.

Let us consider the region around one high energy event with E$_{1}$~$\geq$~10$^{19.8}$~eV. Its coordinates are redefined as (0,0) in Fig.~\ref{method}. First of all, we restrict our analysis to the cosmic rays which angular distance to this event is lower than a given value. This maximal angular distance is denoted by the big black circle in Fig.~\ref{method}. Its radius R depends on the energy threshold E$_{th}$ and has to be optimized for each E$_{th}$ and E$_{1}$. In case the cosmic rays produced by the source are roughly deflected along a line with deflections proportional to 1/E, one can expect that R varies in the following way:

\begin{equation}
\mbox{R}\left(\mbox{E}_{th}\right)= \mbox{R}\left(10^{19.8}\mbox{eV}\right)\frac{10^{19.8}\mbox{eV}}{\mbox{E}_{th}}~,
\label{radius}
\end{equation}

where $\mbox{R}\left(10^{19.8}\mbox{eV}\right)=3-6^\circ$ depending on the position on the sky, the model of GMF and the composition of primaries.
This linear dependence is satisfied for approximately E$_{th}$~$>$~10$^{19.6}$~eV. For lower energies, optimum radii are smaller than radii given by this equation. The optimization of all parameters is discussed in Section~\ref{SectionOptimization}.

If the turbulent field strength B$_{rms0}$ is not too big, most of the lower energy source events are deflected in a region that is sector-shaped (see, for example, Fig.~\ref{BandS}a). A good way to find the direction of this sector is to look for a second highest energy CR around the highest energy one. The line containing both events should generally point at the tail of lower energy events from the source. Let us denote E$_{2}$ (E$_{2}$~$\leq$~E$_{1}$), the energy of the second highest energy event.

In order to discriminate between cases when the highest energy event comes from the source or from the background, one has to set two requirements:
\begin{enumerate}
 \item Its energy E$_{2}$ has to be above a given threshold E$_{2min}$: E$_{2}$~$\geq$~E$_{2min}$. We will show below that when the contribution of the turbulent field is not negligible, one should set E$_{2min}$ to 10$^{19.8}$~eV. Otherwise, one is too often confused by detecting fake sources in the background.
 \item Its angular distance to the highest energy event $d_{12}$ should not exceed a maximum value. We use the following condition:
\begin{equation}
d_{12} \leq \beta \frac{\mathcal{D}}{\mbox{E}_{2}}-\frac{\mathcal{D}}{\mbox{E}_{1}}
\end{equation}
where $\beta \sim 1$ is a parameter that is optimized (see below).
This second requirement guarantees the compatibility with an emission from a same source. If the angular distance is too big, the second highest energy cosmic ray probably comes from the background and one cannot trust it to find the direction of the sector.
\end{enumerate}

Both above conditions are optimized to increase the detection of the source and reduce the background of fake sources.
If more than one event in the circle fulfils these two conditions, then the different potential second highest energy events are tested by decreasing energy order.

The next step is to define the sector. In the following study, we only work with the events located in the sector, assuming that the regular magnetic field dominates deflections. Its central axis is defined as the line which contains the two highest energy cosmic rays. The optimum value of the sector angle $\Theta$ mostly depends on the turbulent field strength. In Fig.~\ref{method}, the sector is highlighted in grey, its central axis is the central black line and the black arrow points at the second highest energy event. The use of the sector enables us to reduce significantly the background. In Section~\ref{results}, we take $\Theta$~=~60$^{\circ}$, which reduces the background by a factor 6.

Then, we rotate from the old coordinates (X,Y) to new ones (X',Y') -cf. Fig.~\ref{method}. The direction of X' is given by the center of mass of all points in the sector, which is not exactly the direction of the central axis of the sector.

The correlation coefficient between X' and 1/E, Corr(X',1/E), can be defined as following~\cite{Amsler:2008zzb}:

\begin{equation}
\mbox{Corr}\left(\mbox{X',1/E}\right) = \frac{\mbox{Cov}\left(\mbox{X',1/E}\right)}{\sqrt{\mbox{Var}\left(\mbox{X'}\right) \mbox{Var}\left(\mbox{1/E}\right)}}
\label{corr_coef}
\end{equation}

where Cov and Var respectively denote covariance and variance:

\begin{equation}
\mbox{Cov}\left(\mbox{X',1/E}\right) = \frac{1}{n} \displaystyle \sum_{i=1}^{n}\left(\mbox{X'}_i - \left\langle \mbox{X'}\right\rangle \right) \left(\mbox{1/E}_i - \left\langle \mbox{1/E}_i\right\rangle \right)
\end{equation}

\begin{equation}
\mbox{Var}\left(\mbox{X'}\right) = \langle \mbox{X'}^2\rangle - \left\langle \mbox{X'} \right\rangle^2 = \frac{1}{n} \displaystyle \sum_{i=1}^{n}\left( \mbox{X'}_i - \left\langle \mbox{X'}\right\rangle \right)^{2}
\end{equation}
if $n$ is the total number of points for which the correlation coefficient is computed.

The next step consists in discriminating the considered selected regions according 
to the value of the correlation coefficient Corr(X',1/E) defined in Eqn.~(\ref{corr_coef}). A minimal value of this coefficient below which the regions are not detected, C$_{min}$, can be found by optimizing the rejection of the background as discussed below.

The use of the correlation coefficient between X' and 1/E was first suggested in Ref.~\cite{Golup:2009zg}, for deflections in the regular field ($\mathcal{R}_{turb/reg}$~$\ll$~1). When the contribution of the turbulent field is added, there are two cases: for values of $\mathcal{R}_{turb/reg}$ lower than $\sim$~0.25, one can still benefit from the correlation coefficient method depicted below, but when $\mathcal{R}_{turb/reg}~\sim$~0.4, Corr(X',1/E) for source events drops to values comparable to those for background events (see Fig.~\ref{CCfor2plus1}).

The next step is to erase in the considered sector the points which decrease Corr(X',1/E). In principle, most of these odd points come from the background.
The CR that are located in the sector are tested one after another, by increasing energy order. We do so because lower energy CR more probably come from the background. Each tested CR is temporarily removed. Then the new axes (X',Y') are redetermined, because the direction of X' is defined by the center of mass of the considered points in the sector, and Corr(X',1/E) is recomputed. If the new value of Corr(X',1/E) is lower than the previous one, the tested point is rewritten in the list of CR and is definitely kept. Otherwise, it is definitely removed. In Fig.~\ref{method}, the events that are kept are surrounded with magenta circles. These tests stop in two cases. First, if Corr(X',1/E)~$\geq$~C$_{min}$. Second, if all the events in the sector have been tested, or if there are only 3 points left because Corr(X',1/E) would not make any sense with less than 3 points.

If the final correlation coefficient is larger or equal to C$_{min}$, the considered selected region around the highest energy event is detected: if it was a source event, the source is detected. If Corr(X',1/E)~$<$~C$_{min}$, and if there is another second highest energy event, the program tries the new associated sector and redoes the depicted method. Otherwise, the region is rejected and there is no detection.

We tested if we would have better results if we only check the points at the border of the sector, those which are far from the central axis, instead all of them. But if $\Theta$ is correctly optimized, no significant difference could be seen on the results.

\subsection{Optimization of the parameters of the method}
\label{SectionOptimization}

\begin{figure}
\centering
\includegraphics[width=0.8\textwidth]{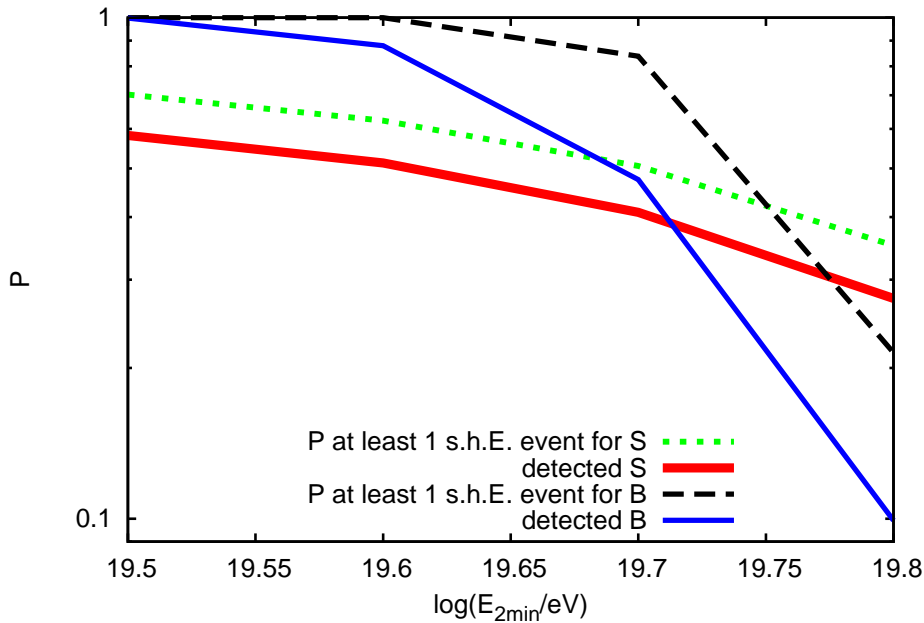}
\caption{Dependence on the minimum energy for the second highest energy event, E$_{2min}$. Green dotted and black dashed lines for the probabilities to have at least one second highest energy event above E$_{2min}$ around a source or a background highest energy event, respectively. Red thick and blue thin solid lines for the probabilities to detect respectively the source and the background through the procedure. Source located in a region of the sky where $\mathcal{D}$~$\simeq$~5.3$^{\circ}$~$\times$~40~EeV and $\mathcal{R}_{turb/reg}~\simeq$~0.2. Source of luminosity equal to 2.6\%, E$_{th}$~=~10$^{19.3}$~eV, 5000 events above 10$^{19}$~eV in the whole visible sky, B$_{reg}$~=~2~$\mu$G and B$_{rms0}$~=~4$\mu$G.}
\label{ShE}
\end{figure}

\begin{figure}
\includegraphics[width=0.5\textwidth]{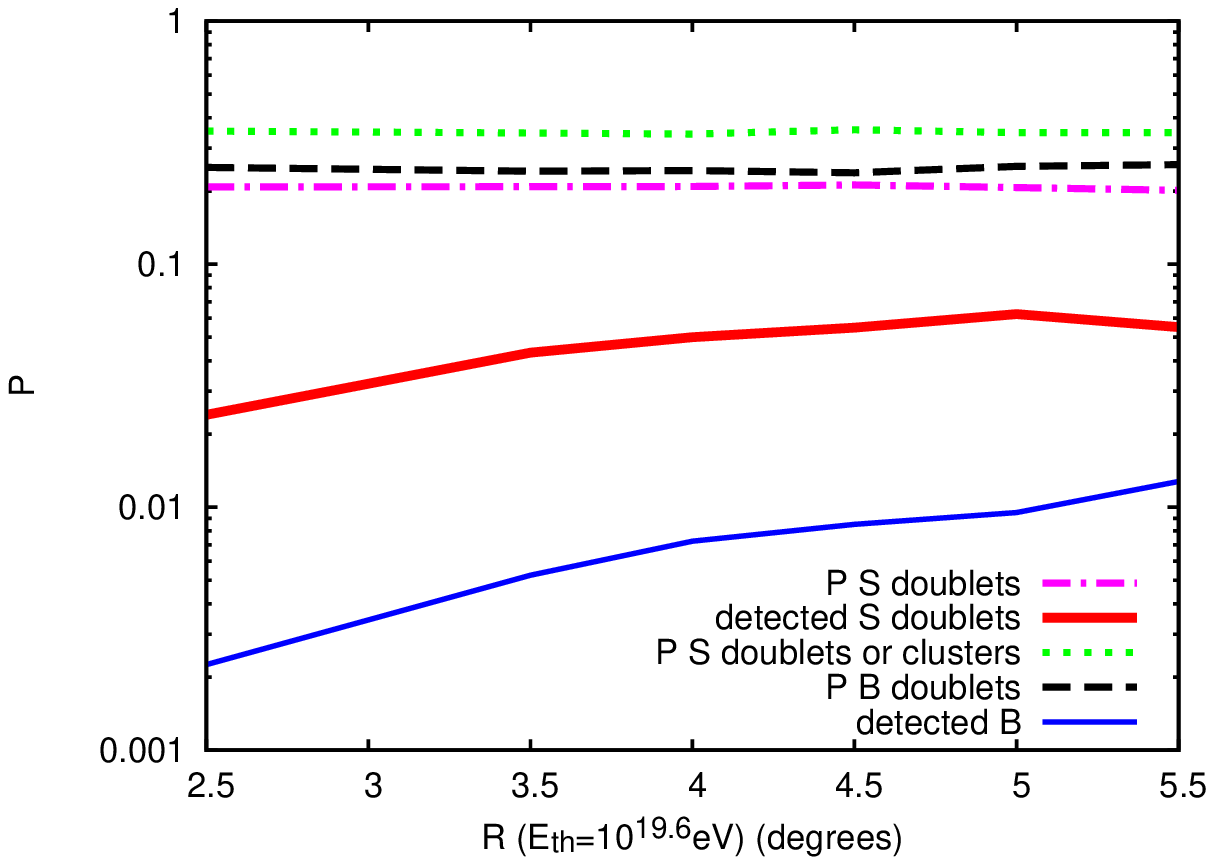}
\includegraphics[width=0.5\textwidth]{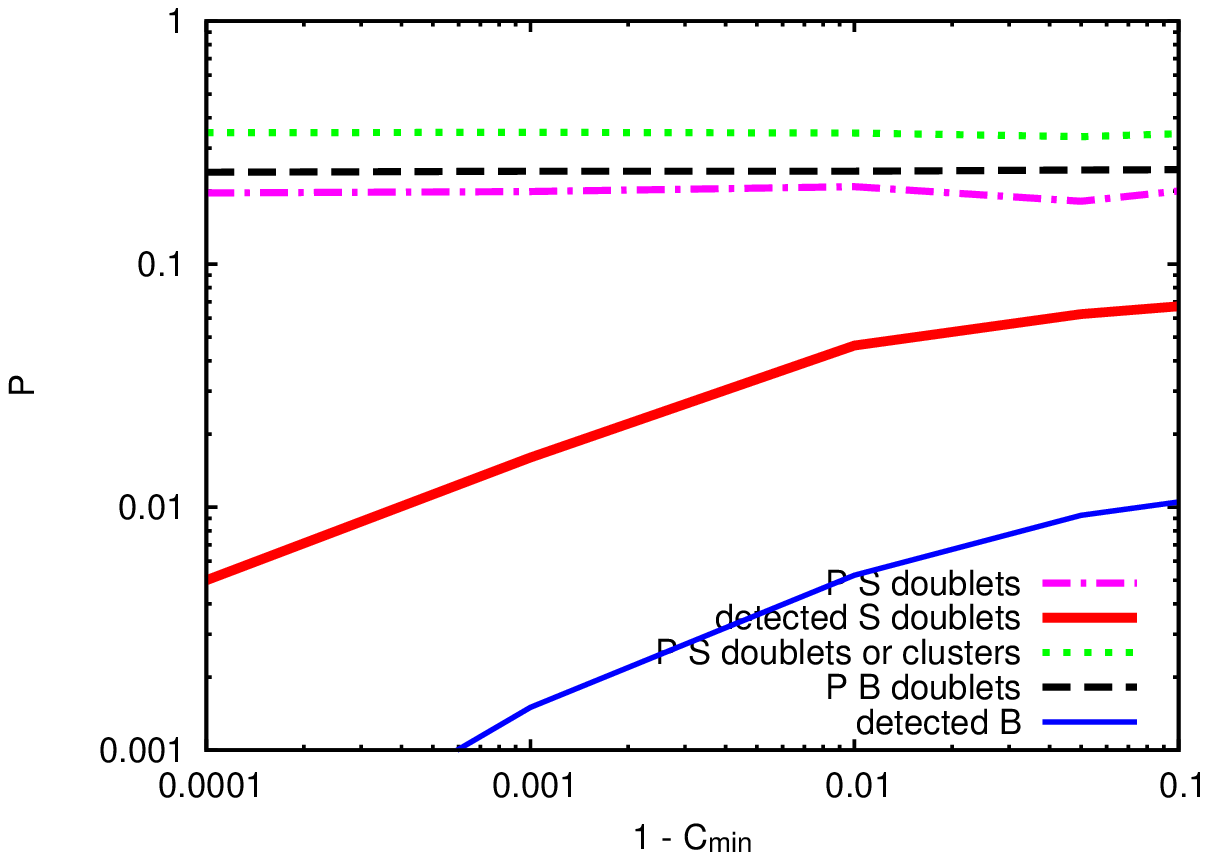}
\includegraphics[width=0.5\textwidth]{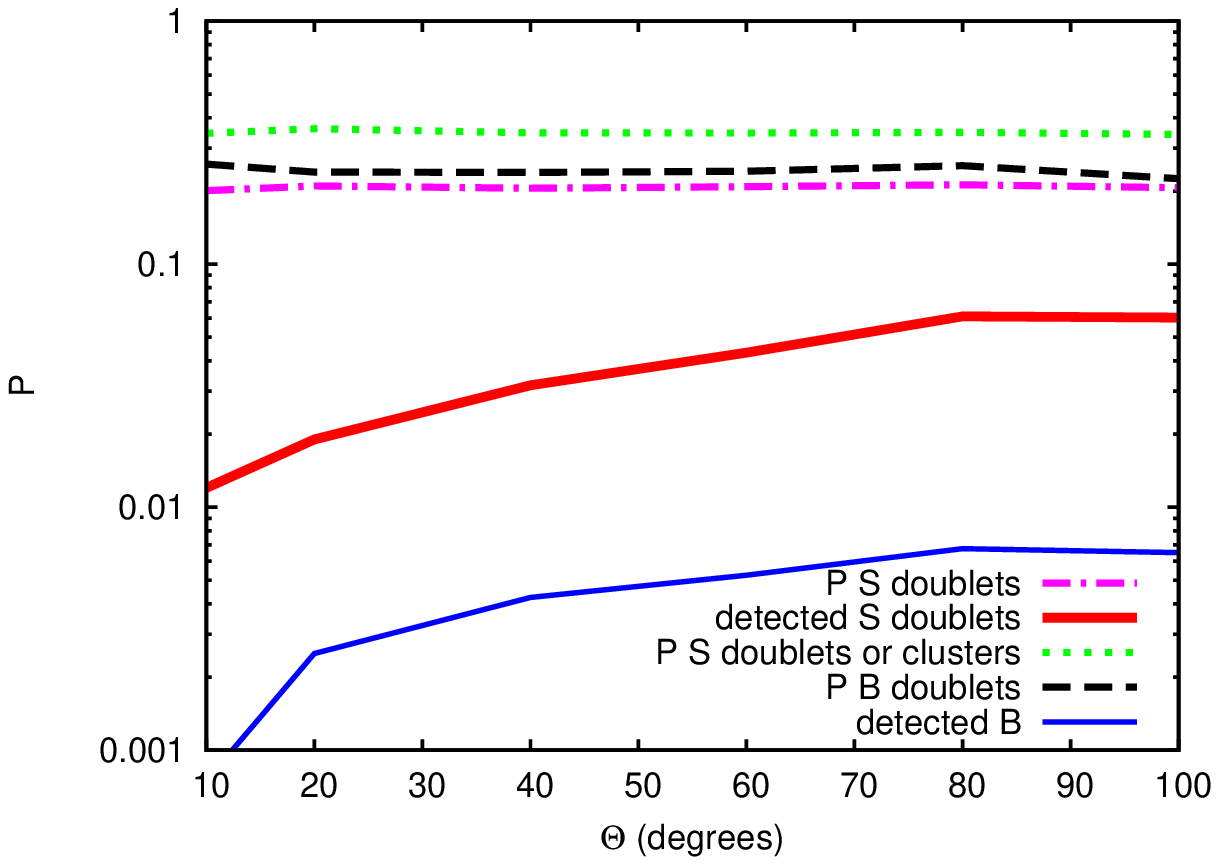}
\includegraphics[width=0.5\textwidth]{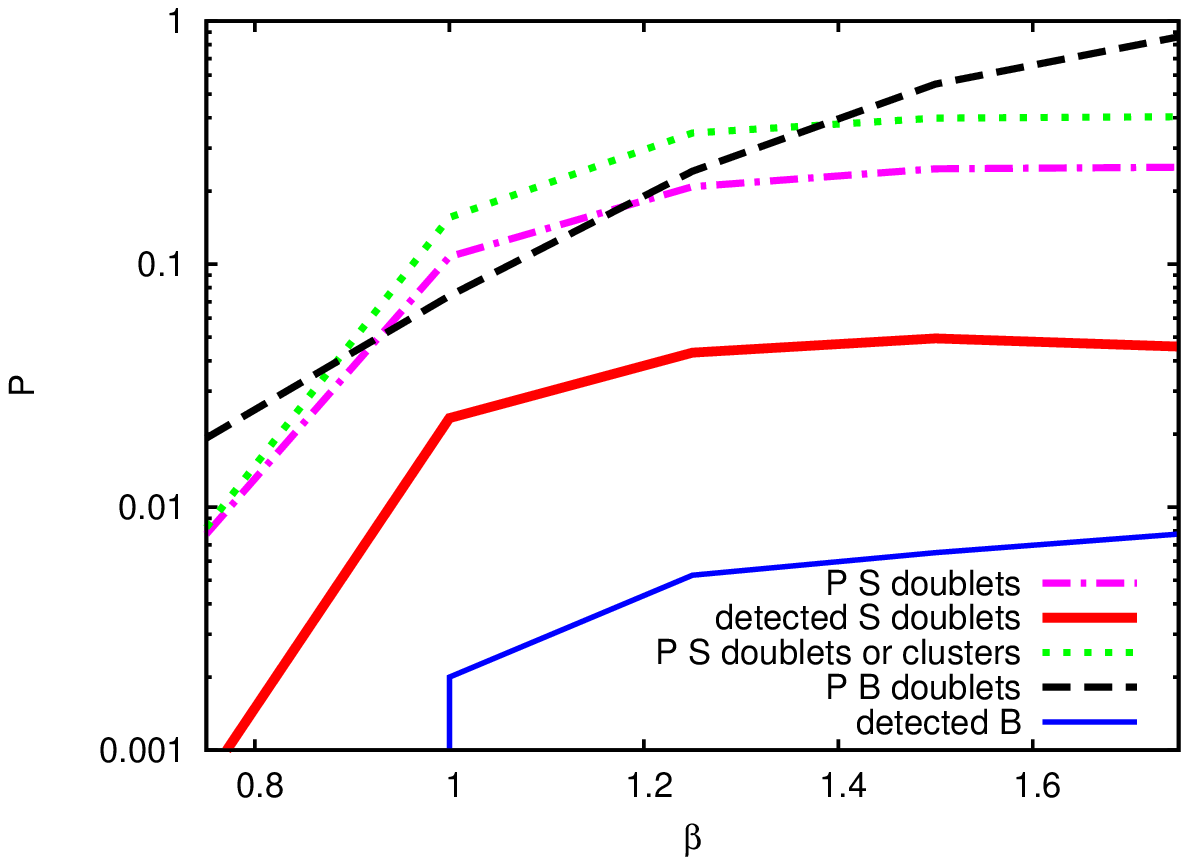}
\caption{Dependence on the four other internal parameters, at E$_{th}$~=~10$^{19.6}$~eV. {\bf Top left corner (a):} radius of the circle; {\bf top right (b):} coefficient C$_{min}$; {\bf bottom left (c):} sector angle $\Theta$; {\bf bottom right (d):} coefficient $\beta$. Magenta dashed-dotted line and red thick solid line for the probabilities to have a source doublet and to also detect it, respectively. Black dashed line and blue thin solid line for the probabilities to have a background doublet and to also detect it, respectively. Green dotted line for source doublets and clusters. Difference between doublets and clusters explained in Section~\ref{results}.}
\label{Optimization}
\end{figure}

In the previous section, we defined five parameters which must be optimized: R, C$_{min}$, $\Theta$, $\beta$ and E$_{2min}$. They are ``internal'' parameters of the method. Their optimum values can depend on physical or experimental parameters, as the regular and turbulent magnetic fields strengths, the energy threshold of sky maps, etc. Optimization is not unique. It is a compromise between the acceptance of the signal and the reduction of the background. Besides, the optimized parameters are not independent.

The optimization of E$_{2min}$, the minimum energy for the second highest energy event, is shown in Fig.~\ref{ShE}. When the turbulent field strength is not negligible, one has to require that E$_{2min}$~=~10$^{19.8}$~eV. If we set E$_{2min}$ to lower values, the probability to detect the source is overwhelmed by the probability to detect some background. The detection of the source tends to become hopeless for E$_{2min}$ below 10$^{19.8}$~eV. Therefore, we will only search for at least doublets of events above 10$^{19.8}$~eV in the sky. In the following, the word ``doublets'' (from the source or from the background) will implicitly refer to such doublets of events with energies above 10$^{19.8}$~eV.

If the turbulent field is negligible as in Ref.~\cite{Golup:2009zg}, one could take a smaller sector angle $\Theta$ and use a lower value of E$_{2min}$.

In Fig.~\ref{Optimization}, we give examples of plots used to optimize the four other parameters. They are done for a source which luminosity is equal to 2.6\% of the total luminosity in UHECR above 10$^{19.8}$~eV. It corresponds to the absolute luminosity L~$\simeq$~1.5$\cdot$10$^{41}$~erg$\cdot$s$^{-1}$ for E~$>$~10$^{19.8}$~eV and for a source located at 100~Mpc from the Earth. We used, as an example, the measured flux and the detector parameters of Auger experiment~\cite{FluxAuger}. We also fixed here E$_{th}~=~10^{19.6}$~eV and B$_{rms0}~=~4~\mu$G.
For this particular example, there is a range of parameters in which the values are close to optimal.

In Fig.~\ref{Optimization}, we can take for example $\beta$~=~1.25, because the probability to have a doublet from the source is already maximal for this value. Taking a higher value would just increase the background. The best value for the opening angle of the sector is in the range $\Theta~\simeq~60^{\circ}-80^{\circ}$.

\subsection{Reconstruction of the source position}

When the source is detected with the method discussed above, the program reconstructs its position, and computes the local deflection power of the regular field in the surrounding region, $\mathcal{D}$. To do so, we plot 1/E versus X' for all the events which are recognized as source events and fit it with a straight line. From this fit, we can compute the deflection power and reconstruct on the axis X' the position of the source -see Fig.~\ref{method}.

%
%
\section{Results}
\label{results}

In this Section, the results are presented for a source which is located in a region where $\mathcal{D}$~$\simeq$~5.3$^{\circ}$~$\times$~10$^{19.6}$~eV $\simeq$~5.3$^{\circ}$~$\times$~40~EeV and $\mathcal{R}_{turb/reg}~\simeq$~0.2, except for Fig.~\ref{RegFieldTurbField}. The value of $\mathcal{R}_{turb/reg}$ is not negligible compared to previous studies -see for example Ref.~\cite{Golup:2009zg} where $\mathcal{R}_{turb/reg}$ is low. But the value $\mathcal{R}_{turb/reg}~\simeq$~0.2 is still significantly below the maximum limit it can reach according to Ref.~\cite{Tinyakov:2004pw}.

In the following, we discuss the dependence of the probability to detect the source on physical parameters of the source and the magnetic field. For the sources which are detected, we find their position on the sky and the local deflection power of the magnetic field.

\subsection{Detection of the source: dependence on parameters}

In the following, we define ``clusters'' as at least three events with energies above 10$^{19.8}$~eV and located in the same region of the sky, according to the condition given by~Eqn.~(\ref{radius}).
For 5000 events above 10$^{19}$~eV in the whole visible sky and no anisotropy, the probability to have a cluster of background events is rare ($\sim$~1\% only), whereas the probability to have doublets of background is of the same order than for source doublets. 
Therefore doublets of events and clusters above 10$^{19.8}$~eV have to be studied separately.
For clusters, the method of Section~\ref{sectionmethod} is not useful because it is optimized to reduce the background which is already low in this case with the price of loosing some real sources. 
On the contrary, for doublets, the method is necessary. It reduces the probability to detect the source too, but it reduces more strongly the probability to detect the background. Therefore we separate doublets and clusters on all plots and mostly focus on doublets in this section. However, for higher numbers of events in the sky, clusters of background events are more frequent and the method of Section~\ref{sectionmethod} can be used for clusters in this case too.

\begin{figure}
\includegraphics[width=0.5\textwidth]{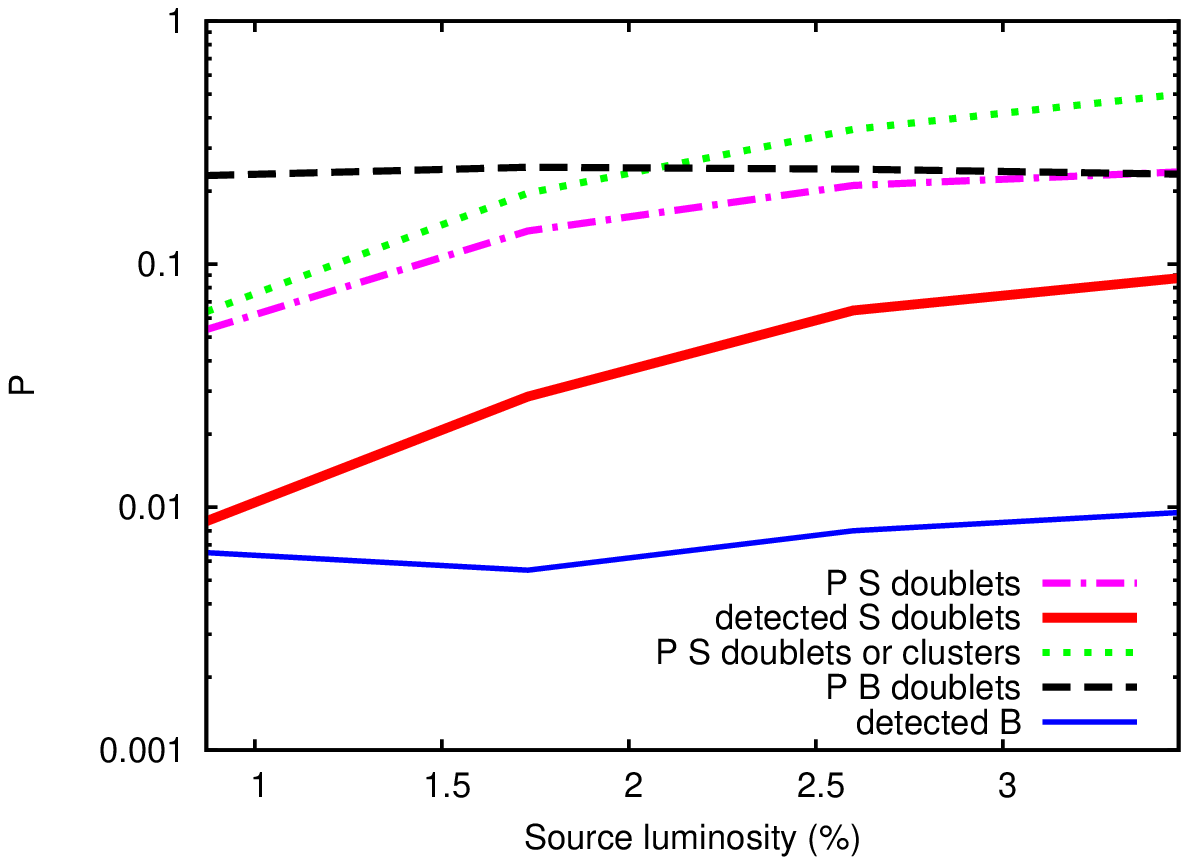}
\includegraphics[width=0.5\textwidth]{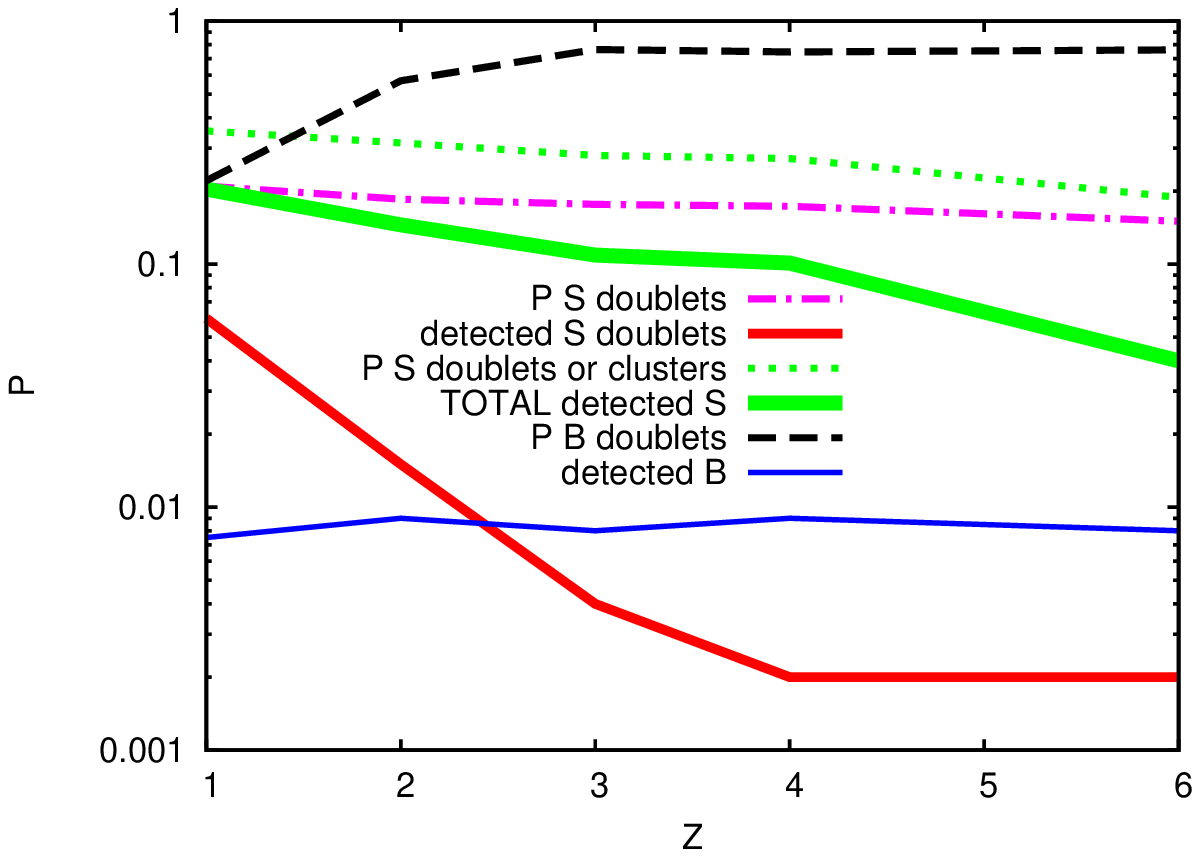}
\caption{Dependence on source parameters. {\bf Left (a):} source luminosity (proton source). {\bf Right (b):} CR composition (2.6\% for source luminosity). Very thick green solid line for the total probability to detect the source. E$_{th}$~=~10$^{19.6}$~eV, 5000 events above 10$^{19}$~eV in the whole visible sky, B$_{reg}$~=~2~$\mu$G and B$_{rms0}$~=~4~$\mu$G.}
\label{LuminosityZ}
\end{figure}

\begin{figure}
\includegraphics[width=0.5\textwidth]{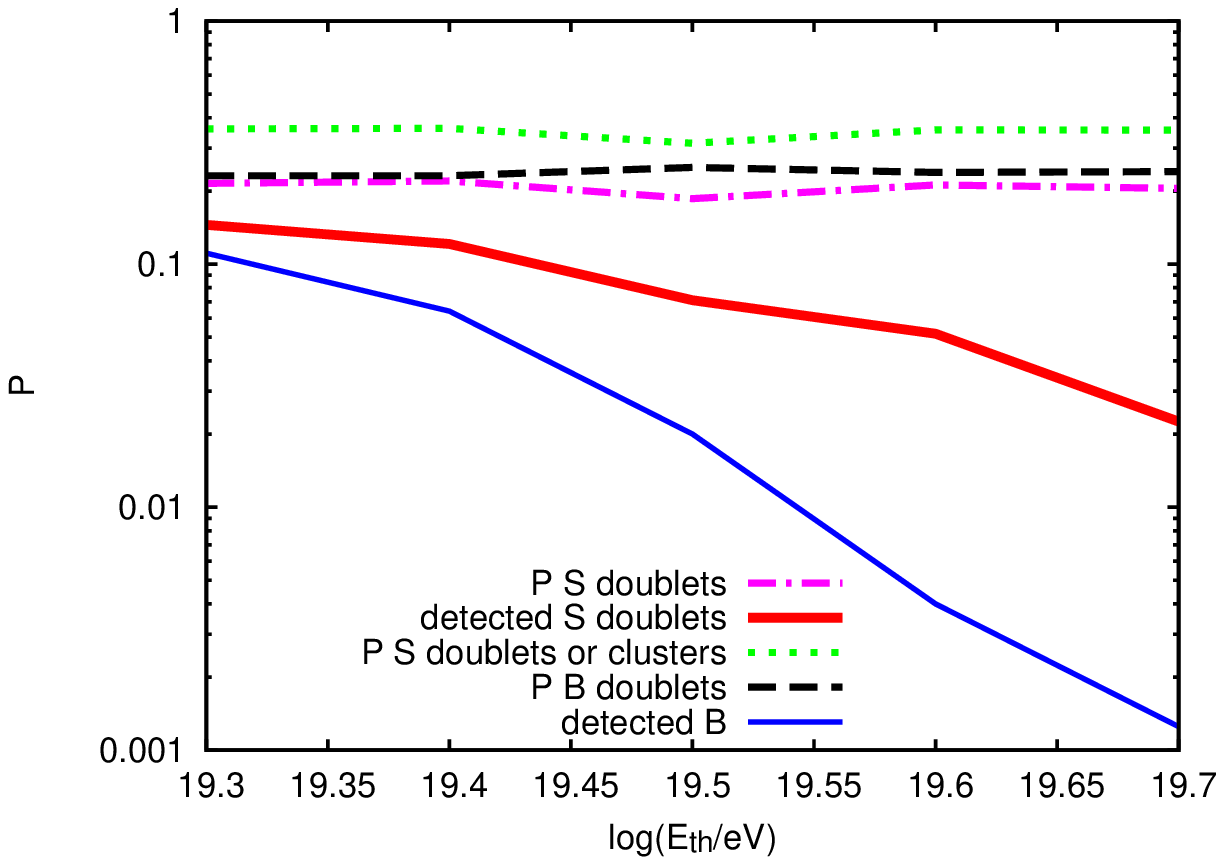}
\includegraphics[width=0.5\textwidth]{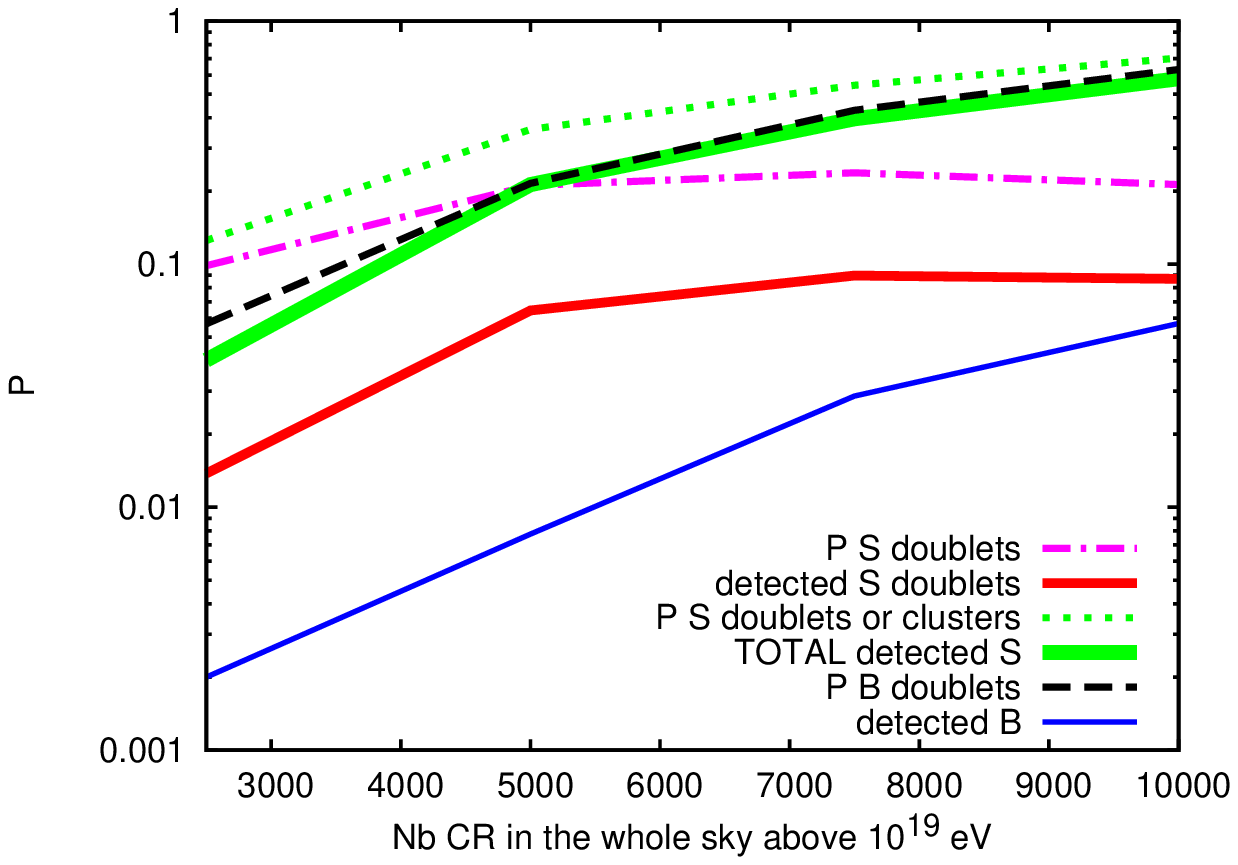}
\caption{Dependence on statistics. {\bf Left (a):} energy threshold of sky maps. {\bf Right (b):} total number of cosmic rays in the sky, above 10$^{19}$~eV. 2.6\% for source luminosity, B$_{reg}$~=~2~$\mu$G and B$_{rms0}$~=~4~$\mu$G. 5000 events above 10$^{19}$~eV in the whole visible sky for (a), and E$_{th}$~=~10$^{19.6}$~eV for (b).}
\label{EthresholdNbCR}
\end{figure}

\begin{figure}
\includegraphics[width=0.5\textwidth]{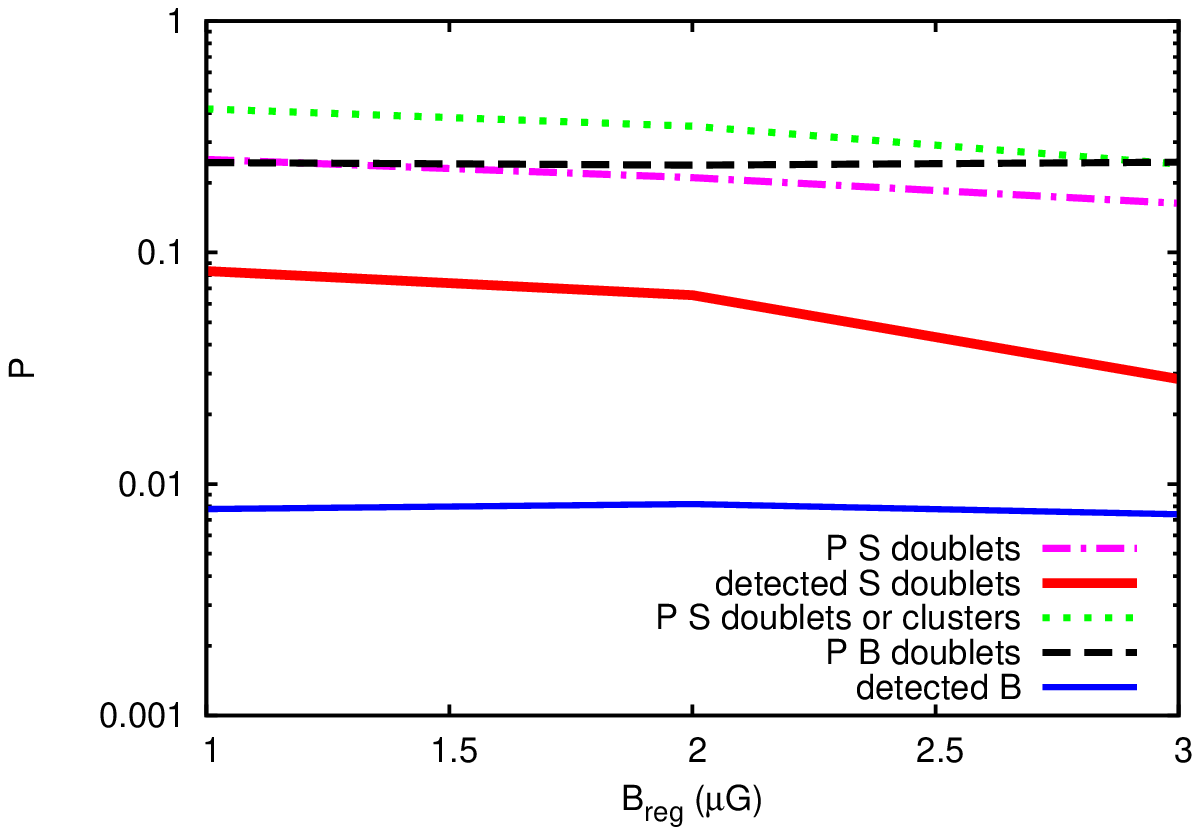}
\includegraphics[width=0.5\textwidth]{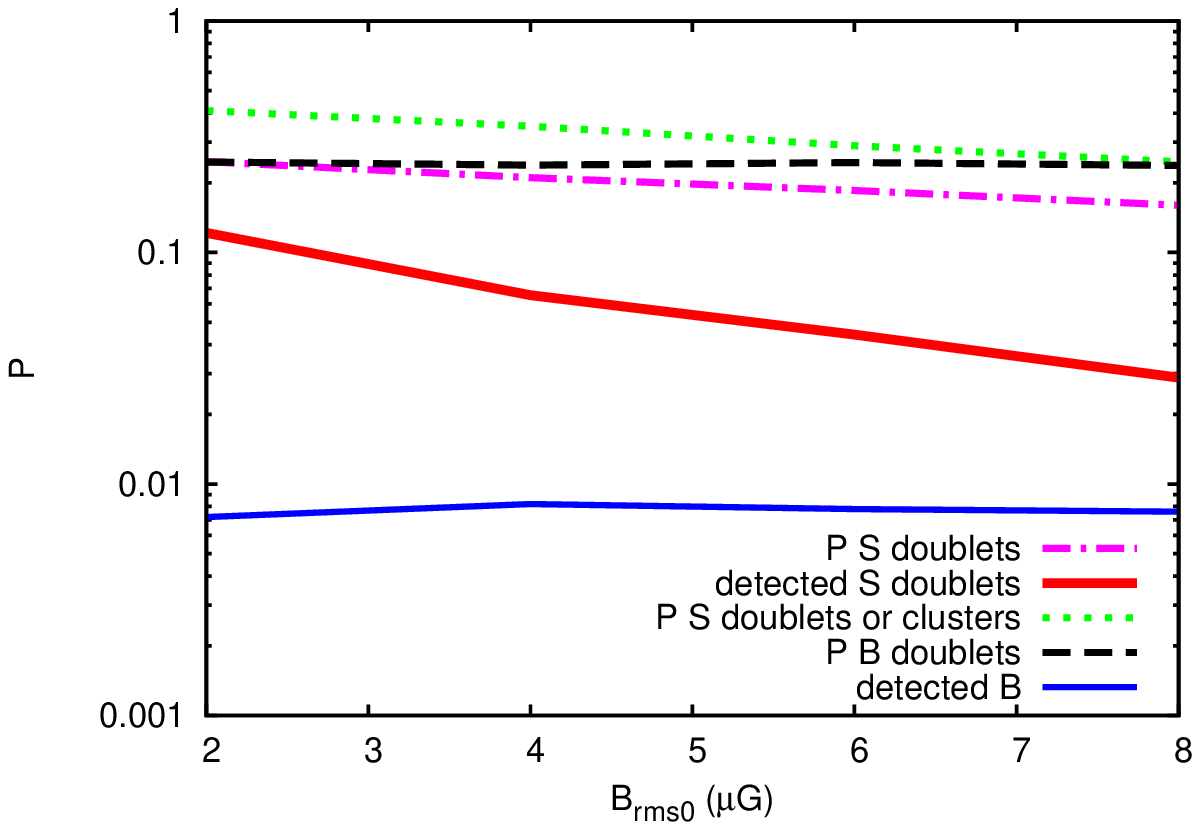}
\caption{Dependence on GMF parameters. {\bf Left (a):} regular field strength close to the Sun, B$_{reg}$, with a constant ratio B$_{reg}$/B$_{rms0}$ set to 0.5. {\bf Right (b):} turbulent field strength, B$_{rms0}$, with B$_{reg}$~=~2~$\mu$G. 2.6\% for source luminosity, E$_{th}$~=~10$^{19.6}$~eV, 5000 events above 10$^{19}$~eV in the whole sky.}
\label{RegFieldTurbField}
\end{figure}

In Figs.~\ref{LuminosityZ}-\ref{RegFieldTurbField}, we plot respectively with the magenta dashed-dotted and the black dashed lines, the probabilities to have on the visible sky a doublet of events from the source or from the background. This first value gives an estimate of how many sources of equivalent luminosities would be required so that one of them would emit a doublet. The red thick and the blue thin solid lines respectively correspond to the probabilities to detect the doublets from the source and from the background with the method proposed in Section~\ref{sectionmethod}. The ratios between these two sets of lines respectively show the efficiencies of source detection and background rejection, when one starts with a high energy doublet. The green dotted lines correspond to the probability to have either a doublet or a cluster from the source. The total probability to detect the source is shown on two plots with very thick green lines. It corresponds to the sum of the probability to detect doublets and of the probability to have a cluster, which has to be seen as a detection in $\sim$~99\% of cases.

The results depend on parameters of the source: its luminosity and the composition of primaries. Fig.~\ref{LuminosityZ}a shows an example of quantitative results for the dependence on the source luminosity, in the range 0.9\% to 3.5\%, for 5000 events above 10$^{19}$~eV in the whole sky and E$_{th}~=~10^{19.6}$~eV. 0.9\% corresponds to a mean value of 0.5 events emitted by the source above 10$^{19.8}$~eV, and 3.5\% to 2 -luminosities of $\sim (0.5-2) \cdot 10^{41}$~erg$\cdot$s$^{-1}$ for sources at 100~Mpc. For a luminosity of 2.6\%, the probabilities to have a doublet or a cluster from the source are similar because the source is bright enough to emit with comparable probabilities 2 or ``3 or more'' events above 10$^{19.8}$~eV. Indeed, for such a number of events in the sky, it corresponds to a source which emits a mean value of about 1.5 events above 10$^{19.8}$~eV.

If the density of UHECR sources is low, we always can consider that their locations on the sky are well separated. For a high density of sources, it is not so, especially in the centers of galaxy clusters. However, since the angular resolution of detectors is 1-2 degrees, all sources located in the center of a galaxy cluster can be considered as an effective single source. Such centers of galaxy clusters would serve as effective bright sources for our study. A very few nearby galaxy clusters are exceptions to this, because they are not point-like -see example of the Virgo cluster in Ref.~\cite{Dolag:2008py}.

In Fig.~\ref{LuminosityZ}b, we investigate the efficiency of the method for sources of light nuclei with Z~$\leq$~6. For Z~$\geq$~3, the background detection is larger than the source detection. $^4$He photo-disintegrates at 10~Mpc distances. Then, in these conditions, light nuclei can only be detected in case of a bright enough source, thanks to clusters of events at high energies -see very thick green solid line. In case of experimental statistics much higher than 5000 events above $10^{19}$ eV, one could take higher energy thresholds, which would improve the detection of light nuclei sources -see below. For low $\mathcal{R}_{turb/reg}$, this method could also be optimized for and applied to bright sources of heavy nuclei in the future.

With such statistics and values, there is no visible difference between the two source spectra ($\alpha$~=~2.2 and $\alpha$~=~2.7). The lower energy tail around the source does not contain enough events to distinguish the power law index.

The results also depend on statistical parameters: the energy threshold of sky maps, E$_{th}$, and the total number of CR detected by the experiment above a given energy which we will set to 10$^{19}$~eV for this discussion. 
Fig.~\ref{EthresholdNbCR}a shows the impact of the energy threshold E$_{th}$ for a proton source, with 5000 events in the whole sky above 10$^{19}$~eV and a luminosity equal to 2.6\%. The higher the energy threshold, the lower the probability to detect the doublets emitted by the source. However, when the energy threshold is decreased, the probability to be confused by the background grows faster than the probability to detect the source. Giving a general optimum energy threshold for source detection is not possible. For example E$_{th}$~=~10$^{19.6}$~eV $\simeq$~4.0~$\cdot$~10$^{19}$~eV (or E$_{th}$~=~10$^{19.5}$~eV $\simeq$~3.2~$\cdot$~10$^{19}$~eV) can be typically good compromises between sufficient source detection and sufficient background rejection. Fig.~\ref{EthresholdNbCR}b is done for the same source, with E$_{th}$~=~10$^{19.6}$~eV and for a total number of cosmic rays in the sky above 10$^{19}$~eV ranging from 2500 to 10000 events, which corresponds to future values for Telescope Array. For such a luminosity, the line corresponding to the detection of the source doublet grows slower than the line corresponding to the detection of background. This is just due to the fact that when the experimental statistics increase, the probability to have a doublet becomes less important for such a bright source and it starts to be overwhelmed by the probability to have a cluster. The total probability to detect the source does not grow slower than the background -see very thick green line.

For one order larger statistics (10000-100000 events), one could increase both E$_{th}$ and the energy required for the two events of the high energy doublet (currently set to 10$^{19.8}$~eV), because the statistics at very high energies would be multiplied by~$\sim$~10. Note that this would improve the efficiency of the method for heavier nuclei.

\begin{figure}
\centering
\includegraphics[width=0.6\textwidth]{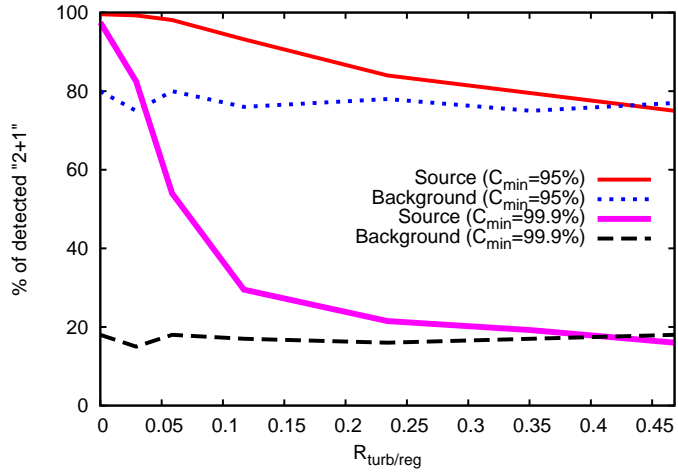}
\caption{Limitations of the correlation coefficient method: dependence on the ratio $\mathcal{R}_{turb/reg}$ of the percentages of detected source and background ``2+1'' -see text. Thin red and thick magenta solid lines for detected ``2+1'' from the source with C$_{min}$~=~0.95 and C$_{min}$~=~0.999, respectively. Dotted blue line and dashed black line for detected ``2+1'' from background with C$_{min}$~=~0.95 and C$_{min}$~=~0.999, respectively. 2.6\% for source luminosity, 5000 events above 10$^{19}$~eV in whole the sky, E$_{th}$~=~10$^{19.6}$~eV and B$_{reg}$~=~2~$\mu$G. $\mathcal{D}$~$\simeq$~5.3$^{\circ}$~$\times$~40~EeV.}
\label{CCfor2plus1}
\end{figure}

We also study the dependence on GMF parameters. In Fig.~\ref{RegFieldTurbField}a, we plot the probability to detect the source versus the regular field strength, with a constant ratio between the regular and the turbulent fields strengths set to B$_{reg}$/B$_{rms0}$~=~0.5. In Fig.~\ref{RegFieldTurbField}b, we show the dependence of the probability to detect the source on the turbulent field strength B$_{rms0}$, with a fixed regular field strength of 2~$\mu$G. The five internal parameters of the method have been optimized for B$_{reg}$~=~2~$\mu$G and B$_{rms0}$~=~4~$\mu$G. They are kept unchanged for the other points of these figures. As expected, for high values of the regular and turbulent fields, the results are worse. However, they are still quite acceptable even for the highest field strengths which are considered in Fig.~\ref{RegFieldTurbField}.

With the parameters discussed above, in 60 \% of cases there is only one event in the energy range 10$^{19.6}$~eV~$<$~E~$<$~10$^{19.8}$~eV. We call such cases "2+1". The fact we require at least three events above E$_{th}$ already allows to discriminate efficiently between source and background doublets. In addition to this requirement, the use of the sector is another key point to detect the source. Note that for the remaining 40\% of cases, the probability to have ``2+more~than~1'' from the background is negligible compared to the probability to have it from the source. The method can also be optimized for stronger magnetic fields by changing the requirement on the minimal number of events in the sector.

For the case of ``2+1'' events, one can see the limitations of the correlation coefficient method for large values of the turbulent GMF. Fig~\ref{CCfor2plus1} gives the fraction of ``2+1'' cases from the source and the background that are detected through the correlation coefficient method, for different ratios $\mathcal{R}_{turb/reg}$, and for two different values of C$_{min}$ (0.95 and 0.999). It shows that one can still strongly improve this discrimination between ``2+1'' from the source and ``2+1'' from the background by using the correlation coefficient method only for low values of $\mathcal{R}_{turb/reg}$ -below $\sim$~0.1. It is still worth for $\sim$~0.25, but useless for much higher values like $\sim$~0.4. In this later case, for such statistics one can simply count events in the sector and require at least one third event between 10$^{19.6}$~eV and 10$^{19.8}$~eV in order to discriminate between source and background.

The previous results are computed for an isotropic sky, even at the highest energies. If we introduce a reasonable anisotropy in the arrival directions of CR above 10$^{19.8}$~eV, by assuming that all of them are located in a given fraction of the sky, ranging from 25\% to 100\%, we see that the results still change in a linear way. For example, if all events above 10$^{19.8}$~eV are located in 25\% of the visible sky, both the probabilities to have a background doublet and to detect it are multiplied by 4.

\subsection{Reconstruction of the source position and of the regular GMF deflection power}
\label{position}

\begin{figure}
\includegraphics[width=0.5\textwidth]{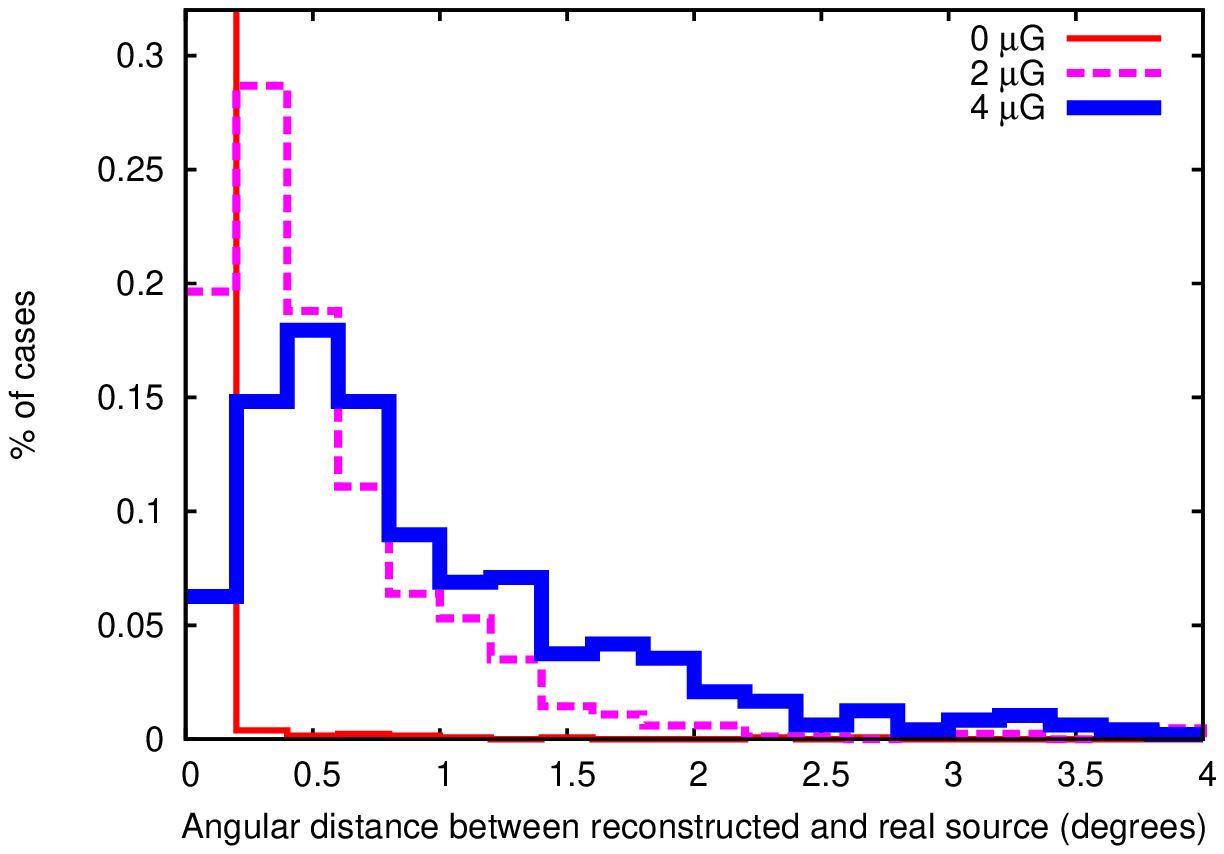}
\includegraphics[width=0.5\textwidth]{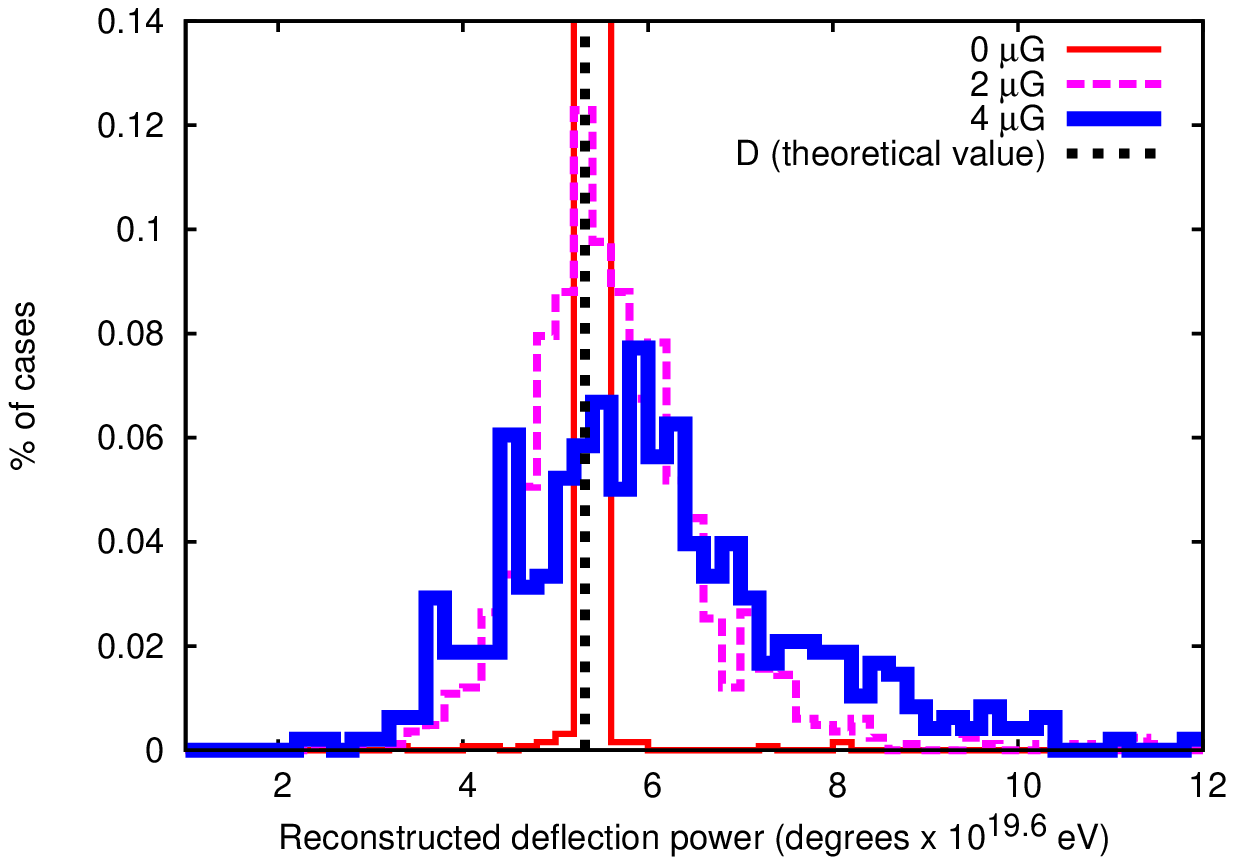}
\caption{{\bf Left (a):} distribution of distances, in degrees, between reconstructed and real sources. {\bf Right (b):} distribution of reconstructed deflection powers in degrees~$\times$~10$^{19.6}$~eV. Thin red solid line for the turbulent component off (0~$\mu$G), magenta dashed line for B$_{rms0}$~=~2~$\mu$G, and thick blue solid line for B$_{rms0}$~=~4~$\mu$G.}
\label{DisdD}
\end{figure}

With a turbulent GMF strength set to B$_{rms0}$~=~4~$\mu$G, the method can reconstruct the source position at less than 1~degree from the real one in 68\% of cases -see the distribution with a blue thick solid line in Fig.~\ref{DisdD}a. This result is satisfying because the experimental angular resolution is close to 1~degree.

Also, for this turbulent field value, the deflection power can be computed with a precision of about $\pm$~25\%, in 68\% of cases -see blue thick solid line in Fig.~\ref{DisdD}b. 68\% of all reconstructed values are between 4.5 and 7.4$^{\circ}\times10^{19.6}$eV, and 95\% are between 3.3 and 12.4$^{\circ}\times10^{19.6}$eV. The distribution has a bigger tail for higher values of $\mathcal{D}$.

The precisions on both results are mostly affected by the turbulent GMF strength: the distributions computed for B$_{rms0}$~=~0~$\mu$G and B$_{rms0}$~=~2~$\mu$G are shown on the same figures for comparison. The lower the turbulent component, the more precise the results. With B$_{rms0}$~=~0~$\mu$G, in most of the cases, the distance between the real and the reconstructed source is much lower than one degree -see thin red lines in Figs.~\ref{DisdD}a and b. Then, the precision of reconstruction would be completely dominated by the experimental angular resolution. The distribution for the reconstructed deflection power $\mathcal{D}$ has a small width and is peaked around the theoretical value.

The value of the regular component -for B$_{reg}$~=~1 to 3~$\mu$G- does not affect the precision on $\mathcal{D}$. For low B$_{reg}$, the results for source reconstruction are more precise.

For the values we consider, the detector resolution on energies does not really affect, statistically, the reconstruction of the source position. If we take $\Delta$E/E~=~25\% which is the order of commonly admitted values, both for the distance between the reconstructed and the real source and for the reconstructed deflection power, no significant difference with $\Delta$E/E~=~0 is found.

No really noticeable impact of the luminosity -between 1.5 and 3.5\%- or the total number of CR in the sky -between 2500 and 10000 above 10$^{19.0}$~eV- could be detected on the results.

\section{Conclusions}
\label{summary}

We investigated here the possibility to identify single bright ultra-high energy cosmic ray sources on the sky if the regular component of the Galactic Magnetic Field gives the major contribution to deflections of UHECR.

To simulate the background, we took events following the Telescope Array exposure. Their energies were distributed according to the cosmic ray spectrum measured by HiRes. We considered the cases of 2500 to 100000 events
with E~$>$~10$^{19}$~eV.

For the single source, we took a power law acceleration spectrum proportional to 1/E$^{2.2}$, with a maximum energy E$_{max}$~=~10$^{21}$~eV. Then, we propagated it from the source, put at distances equal to 50-100~Mpc, to the observer by taking into account all energy losses as in Ref.~\cite{Kachelriess:2007bp}. In order to simplify the study, we assumed, following Ref.~\cite{Dolag:2004kp}, that deflections in the extragalactic magnetic field are small.
We deflected the cosmic rays produced by the source in the Galactic Magnetic Field. We considered a regular component in the disk and the halo similar to the Prouza and Smida model~\cite{PS}, but we modified some parameters according to a recent discussion during the Ringberg workshop on UHECR and magnetic fields~\cite{Ringberg} (for details, see Section~\ref{source}). We also added the turbulent component both in the disk 
and in the halo. Finally, we modified the signal from the source according to the exposure of Telescope Array.

We combined the signals from the source and the background in a common data set and tried to recover the source 
from these data, using the method presented in Section~\ref{method}. As we showed in Fig.~\ref{ShE}, in order to discriminate between the source and the background when the contribution of the turbulent GMF is not negligible, one needs at least 2 events from the source at the highest energies, E~$>$~60~EeV. Then, one can look for the low energy tail, as discussed in Section~\ref{method}.

We studied in Section~\ref{results} the dependence of these results on the main parameters of the source and of the Galactic Magnetic Field.

Finally, we reconstructed the position of the source and the deflection power of the magnetic field 
in Section~\ref{position}. The main result is that the position of the source can be 
reconstructed within one degree from the real position in 68\% of cases. The regular GMF deflection power 
is known with a 25\% precision level for a turbulent field strength equal to 4~$\mu$G close to the Sun. For lower values of the turbulent magnetic field, the precision is much better.

\section{Acknowledgments}

We thank M.Kachelriess and G.Sigl for useful discussions and comments on this work.



\begin{thebibliography}{10}
\bibitem{Ringberg}
Recently, a joint discussion on UHECR and galactic magnetic fields was done in Ringberg Castle. Since there is no Standard Model of the Milky Way magnetic field, we adopt in this paper the parameters discussed on this workshop. In particular, we choose local regular and turbulent fields with values equal to 2 $\mu$G  and 4 $\mu$G, respectively.
\end{thebibliography}
\end{document}